\documentclass[12pt]{iopart}
\usepackage[dvips]{graphicx}%
\usepackage{bm}%
\usepackage{cite}
\usepackage{multirow}

\usepackage{iopams}
\expandafter\let\csname equation*\endcsname\relax
\expandafter\let\csname endequation*\endcsname\relax
\usepackage[fleqn]{amsmath}
\usepackage{etoolbox}
\usepackage[normalem]{ulem} 
\usepackage[colorlinks=true,linkcolor=blue,citecolor=blue ,urlcolor=blue]{hyperref}

\newcommand{\op}[1]{{\bf #1}}
\newcommand{\ket}{\rangle}
\newcommand{\bra}{\langle}

\begin{document}

\title{Storing Quantum Information in a generalised Dicke Model via a Simple Rotation}

\author{R. L\'opez-Pe\~na, S. Cordero, E. Nahmad-Achar, O. Casta\~nos}
\address{Instituto de Ciencias Nucleares, Universidad Nacional Aut\'onoma de M\'exico}

\begin{abstract}
A method for storing quantum information is presented for $3$-level atomic systems interacting dipolarly with a single radiation field.
The method involves performing simple local SU(2) rotations on the Hamiltonian. Under equal detuning, these transformations decouple one of the atomic levels from the electromagnetic field for the $\Lambda$- and $V$-configurations, yielding two effective $2$-level systems (qubits) plus an isolated atomic level; this allows for the exchange of information between the qubits. This rotation preserves the quantum phase diagram of the system. The method could possibly be used as a means to manipulate quantum information, such as storage and retrieval, or communication via a transmission line.
\end{abstract}

\section{Introduction}

The models describing the interaction between an electromagnetic field (EM) and matter  have been very important in quantum optics to discover new phenomena, as well as serving as a  testing arena for  quantum information theory.  Precursors of these models are the Rabi semi-classical model~\cite{rabi37}, and the Dicke model~\cite{dicke54}.

A review of the dynamical features of an atom, with two or three levels, interacting with quantised electromagnetic (QEM) cavity fields in the rotating wave aproximation (RWA) is presented in~\cite{yoo85}, where the Jaynes-Cummings generalization to three levels is established.
In~\cite{li87,liu87,lin89,li89} a more detailed investigation for this three level model interacting with one-mode or two-modes QEM cavity fields was carried out: the dynamic behaviour of level population, coherence of the radiation fields, photon-number distribution, and the dipole moments were calculated.
A generalization of the model to contemplate a finite number of atoms is used in~\cite{cordero13a,cordero13b}, considering the interaction with one-mode QEM field; there, the ground state is calculated using coherent variational states to determine the quantum phase transitions of the system through the expectation value of the   Hamiltonian of the system, and to find the locci of points where a sudden change in the ground state takes place, when the dipolar strengths are varied. 
The interaction between a three-level $\Lambda$-type atom and one-mode electromagnetic cavity field in the presence of the classical homogenous gravitational field has been studied in~\cite{el-wahab15}.
When a time-varying coupling and power-law potentials are allowed in the model, it is shown that a large nonlocal correlation can be achieved through the control of physical parameters~\cite{algarni22}.

On the other hand, there have been efforts to solve the model analytically. In~\cite{puri88} the procedure of adiabatic elimination with one-mode was applied to the $\Xi$-configuration, and an effective two-level Hamiltonian with intensity dependent Stark shift terms was obtained. In~\cite{gerry90} the adiabatic elimination procedure was carried out in the $\Lambda$-configuration with two-modes, obtaining again an effective two-level Hamiltonian with intensity dependent Stark shift terms. In~\cite{alexanian95} it was shown that a unitary operator can be used to transform the previous Hamiltonian to obtain the same results if terms up to second order are kept. This unitary transformation is shown to exactly reduce the system to a two-level one with an effective Raman coupling, which depends nonlinearly on the intensity of the two radiation fields~\cite{wu96}.

In this paper, we consider a generalised Dicke model which consists of a system of $3$-level atoms interacting dipolarly with a single mode of an electromagnetic field. In the atomic scenario, this system has three different configurations, namely $\Xi$ $\Lambda$ and $V$~\cite{yoo85}.  Studies of the properties of the ground state as functions of the control parameters, viz. dipolar strengths, have been considered previously~\cite{cordero13a,cordero13b}. Here, we stress the mathematical structure of the model in the three atomic configurations: the matter contribution is represented by a ${\rm U}(3)$ group structure, and for each atomic configuration a linear combination of the matter operators allows us to construct a natural $\mathfrak{su}(2)$ subalgebra.  

In particular, there are two ${\rm SU}(2)$ local rotations which decouple the EM dipolar interaction between two of the atomic levels, leading to an effective Hamiltonian consisting of the contribution of a $2$-level Dicke model plus an isolated atomic level, and a remnant term corresponding to a one-body matter interaction between the atomic levels with a forbidden dipolar transition. 

We find, for the $\Lambda$- and $V$-configurations, that the one-body matter interaction disappears when two of the atomic energy levels are degenerate (equal detuning). In this case, the expectation value of the population of one or the other atomic level, becomes a constant of motion (which vanishes for the ground state).

For a finite quantum system, with a fixed number $N_a$ of atoms, we use the fidelity and the susceptibility of the fidelity between neighbouring states to determine the quantum phase diagram~\cite{zanardi06}. These detect any abrupt change in the state of a system when a control parameter is varied. By definition, when a unitary transformation is performed on the system, independent of the control parameters, these quantities remain invariant. Equivalently, the phase diagram does not change.

For the particular local SU$(2)$ rotations considered here, which decouple an atomic state, even when they depend on the control parameters $\mu_{ij}$ the minima of the fidelity remain at the same loci as those for the original Hamiltonian, thus maintaining the quantum phase diagram invariant (cf.~\ref{AppB}).

This paper is organised as follows: In Section II the generalised Dicke model is established. Sec. III presents the $\mathfrak{su}(2)$ structure which arises naturally in each configuration, and shows the particular rotation that decouples a level in every case. Sec. IV gives  the expectation values of the atomic energy levels, as well as the quantum phase diagrams.  In the final Section some conclusions and possible applications of the results are presented.

\section{Generalised Dicke Model}

 Let us consider a generalised Dicke model for a system of $N_a$ atoms of $3$-levels interacting dipolarly with a single mode of electromagnetic field, the Hamiltonian of this system reads as ($\hbar =1$)
   \begin{equation}
      \op{H}=\Omega\,\op{a}^{\dagger}\op{a}+\sum_{j=1}^{3}\omega_{j}\,\op{A}_{jj}
      -\frac{1}{\sqrt{N_{a}}}\left(\op{a}^{\dagger}+\op{a}\right)\,
      \sum_{j<k}^{3}\mu_{jk}\,\left(\op{A}_{jk}+\op{A}_{kj}\right)\ ,
      \label{ham3}
   \end{equation}
where $\Omega$ denotes the frequency of the photons, $\omega_k$ the frequency of the atomic energy levels, and $\mu_{jk}$ the dipolar strength of the atomic transition $\omega_j\rightleftharpoons \omega_k$ (used as control parameters of the system). The field creation and annihilation boson operators are denoted by  $\op{a}^\dag$  and  $\op{a}$, respectively. The matter operators by $\op{A}_{jk}$, with bosonic representation $\op{A}_{jk}=\op{b}_j^\dag\op{b}_k$,  annihilate one atom in the atomic level $\omega_k$ and create one atom in the atomic level $\omega_j$; these operators obey an $\mathfrak{su}(3)$ algebra
   \begin{equation}
      \Big[\op{A}_{jk},\,\op{A}_{\ell m}\Big]=\delta_{\ell k}\,\op{A}_{jm}
      -\delta_{jm}\,\op{A}_{\ell k}\ .
   \end{equation}

The Hamiltonian in eq.~(\ref{ham3}) is given in general form. In order to establish a criterion which permits to write the Hamiltonian for each atomic configuration, we adopt the convention $\omega_1\leq\omega_2\leq\omega_3$;  furthermore, we set the field mode frequency  in the cavity to $\Omega=1$, so that  both energy and dipolar strengths are measured in units of $\hbar\Omega$. The nature of the matter-field dipolar interaction requires, by symmetry considerations, that at least one of the coupling constants $\mu_{jk}$ be null, and this gives a forbidden dipolar transition. Using our convention, each atomic configuration is then given by the appropriate vanishing of a dipolar strength, known in the literature as $\Xi$ ($\mu_{13}=0$),  $V$ ($\mu_{23}=0$) and $\Lambda$ ($\mu_{12}=0$) for their schematic representation (cf. Figure~\ref{confXLVfig}).

The corresponding  Hamiltonians preserve the parity of the number of excitations, i.e., $[\op{H},\op{\Pi}]=0$ with $\op{\Pi}=e^{i\pi\op{M}}$ the parity operator and $\op{M}$ the total number of excitations operator, which takes the form  $\op{M}_\Xi =  \op{a}^\dag\op{a} + \op{A}_{22}+2\op{A}_{33}$, $\op{M}_V =  \op{a}^\dag\op{a} + \op{A}_{22}+\op{A}_{33}$ and $\op{M}_\Lambda =  \op{a}^\dag\op{a} +\op{A}_{33}$. These are constants of motion when the rotating wave approximation (RWA) is considered; however in the general case, the parity operators divide the Hilbert space into two subspaces $\op{H}=\op{H}_e\oplus\op{H}_o$ (even and odd total number of excitations).

It is convenient to introduce the detuning for the transitions $\omega_j\rightleftharpoons \omega_k$ with $j<k$,  defined as
\begin{equation}\label{eq.detuning}
\Delta_{jk} :=\Omega-\omega_{jk}\,,\qquad \omega_{jk}:= |\omega_j-\omega_k|\,;
\end{equation}
when two of the atomic energy levels in the $\Lambda$ and $V$ configurations are degenerate, both transitions will have the same detuning, that is, $\Delta_{13}=\Delta_{23}$ and $\Delta_{12}=\Delta_{13}$.

In the next section we explore the effect of simple rotations on the system to obtain new Hamiltonians. We will show that we can always obtain a new system in which the interaction between
a pair of levels with the EM field in the new Hamiltonian is eliminated, leaving an
isolated atomic level and an effective $2$-level Dicke model.

\section{Model Hamiltonians}

It is well known that there are different ${\rm SU}(2)$ groups which can be constructed
from $\mathfrak{su}(3)$ generators. Particularly interesting in the model Hamiltonian~(\ref{ham3})
are the ones that can be formed by adding, to the $\mathfrak{su}(3)$ generators which appear in a given configuration, the missing one to complete an $\mathfrak{su}(2)$ algebra (cf. Table~(\ref{t1})).

\begin{table}[h]
   \caption{$\mathfrak{su}(2)$ subalgebras for the 3-level configurations. }
   \begin{center}
   \begin{tabular}{c||c}
      Configuration & $\mathfrak{su}(2)$ subalgebra generators\\
      \hline\\[-5mm]
      \hline
      \rule[-4mm]{0mm}{11mm}
      $\Xi$ & $\left\{\op{A}_{12}+\op{A}_{21},\,\op{A}_{23}+\op{A}_{32},\,
      i\,\left(\op{A}_{31}-\op{A}_{13}\right)\right\}$
      \\
      \hline
      \rule[-4mm]{0mm}{11mm}
      $\Lambda$ & $\left\{\op{A}_{23}+\op{A}_{32},\,\op{A}_{13}+\op{A}_{31},\,
      i\,\left(\op{A}_{12}-\op{A}_{21}\right)\right\}$
      \\
      \hline
      \rule[-4mm]{0mm}{11mm}
      $V$ & $\left\{\op{A}_{12}+\op{A}_{21},\,\op{A}_{13}+\op{A}_{31},\,
      i\,\left(\op{A}_{32}-\op{A}_{23}\right)\right\}$
   \end{tabular}
   \end{center}
\label{t1}
\end{table}
\noindent
By means of  a rotation   through the third element in each corresponding subalgebra,
   \begin{equation}
     \op{U}_{jk}(\alpha):=\exp\left[-\alpha\, \op{K}_{jk} \right]\ ,
     \label{su3rotation}
   \end{equation}
with
\begin{equation}\label{eq.Kjk}
\op{K}_{31}=\op{A}_{31} - \op{A}_{13},\ \op{K}_{12}=\op{A}_{12} - \op{A}_{21},\ \op{K}_{32}=\op{A}_{32} - \op{A}_{23}\,,
\end{equation}
we see  that, in order to eliminate the interaction between one pair of atomic levels with the EM field in the cavity, one may select the parameter $\alpha$, as follows (cf.~\ref{AppA}):
\begin{enumerate}
\item
$\alpha = \arctan(\mu_{23}/\mu_{12})$ and $\alpha = -\arctan(\mu_{12}/\mu_{23})$ make the interaction of levels $\omega_2\leftrightharpoons\omega_3$ and $\omega_1\leftrightharpoons\omega_2$ with the field vanish respectively in the $\Xi$ configuration;
\item
$\alpha = \arctan(\mu_{13}/\mu_{23})$ and $\alpha = -\arctan(\mu_{23}/\mu_{13})$ make the interaction of levels $\omega_1\leftrightharpoons\omega_3$ and $\omega_2\leftrightharpoons\omega_3$ with the field vanish respectively in the $\Lambda$ configuration;
\item
$\alpha = \arctan(\mu_{13}/\mu_{12})$ and $\alpha = -\arctan(\mu_{12}/\mu_{13})$ make the interaction of levels $\omega_1\leftrightharpoons\omega_3$ and $\omega_1\leftrightharpoons\omega_2$ with the field vanish respectively in the $V$ configuration.
\end{enumerate}

The transformed Hamiltonian may, in general, be written in the form
	\begin{eqnarray}
	   &&\op{H}^{\prime}=\op{U}_{jk}(\alpha)\,\op{H}\,\op{U}_{jk}^{\dagger}(\alpha)
	   \nonumber\\
	   &&=\Omega\,\op{a}^{\dagger}\op{a}+\sum_{\ell=1}^{3}\tilde{\omega}_{\ell}\,\op{A}_{\ell\ell}
	   +\tilde{\lambda}_{jk}\,\left(\op{A}_{jk}+\op{A}_{kj}\right)
	   -\frac{1}{\sqrt{N_{a}}}\left(\op{a}^{\dagger}+\op{a}\right)\,
       \sum_{\ell<m}^{3}\tilde{\mu}_{\ell m}\,\left(\op{A}_{\ell m}+\op{A}_{m\ell}\right)\, .
       \label{ham3p}
	\end{eqnarray}
where the expressions for $\tilde{\omega}_{\ell}$, $\tilde{\lambda}_{jk}$, and $\tilde{\mu}_{\ell m}$ are given in Table~\ref{t2a}. The subscripts on $\tilde{\lambda}$ correspond to the  forbidden dipolar transitions in each atomic configuration, viz., $\tilde{\lambda}_{13}$ for $\Xi$-configuration, $\tilde{\lambda}_{12}$ for
$\Lambda$-configuration, and $\tilde{\lambda}_{23}$ for $V$-configuration.

\begin{table}[h!]
   \caption{Values of the parameters $\tilde{\omega}_{\ell}$, $\tilde{\lambda}_{jk}$, and $\tilde{\mu}_{\ell m}$
   in the Hamiltonian~(\ref{ham3p}), when $\alpha$ is chosen to remove one of the atom-EM field interactions. The subscripts on $\tilde{\lambda}$ correspond to the  forbidden dipolar transitions in each atomic configuration.}
   \vspace{-0.2in}
   \begin{center}
   \begin{tabular}{c||c|c||c|c||c|c}
   Conf.&\multicolumn{2}{|c||}{$\Xi$}&\multicolumn{2}{|c||}{$\Lambda$}&\multicolumn{2}{|c}{$V$}\\
   \hline\\[-5mm]
   \hline
   \rule[-4mm]{0mm}{11mm}
   $\tan(\alpha)$ &$\frac{\mu_{23}}{\mu_{12}}$ &-$\frac{\mu_{12}}{\mu_{23}}$ &$\frac{\mu_{13}}{\mu_{23}}$ 
   &-$\frac{\mu_{23}}{\mu_{13}}$ &$\frac{\mu_{13}}{\mu_{12}}$ &-$\frac{\mu_{12}}{\mu_{13}}$\\
   \hline
   \rule[-4mm]{0mm}{11mm}
   $\tilde{\omega}_{1}$ &$\frac{\omega_{1}\mu_{12}^2+\omega_{3}\mu_{23}^2}{\mu_{12}^{2}+\mu_{23}^2}$
   &$\frac{\omega_{1}\mu_{23}^2+\omega_{3}\mu_{12}^2}{\mu_{12}^{2}+\mu_{23}^2}$
   &$\frac{\omega_{1}\mu_{23}^2+\omega_{2}\mu_{13}^2}{\mu_{13}^2+\mu_{23}^2}$
   &$\frac{\omega_{1}\mu_{13}^2+\omega_{2}\mu_{23}^2}{\mu_{13}^2+\mu_{23}^2}$
   &{\scriptsize$\omega_{1}$} &{\scriptsize$\omega_{1}$}\\
   \hline
   \rule[-4mm]{0mm}{11mm}
   $\tilde{\omega}_{2}$ &{\scriptsize$\omega_{2}$} &{\scriptsize$\omega_{2}$} &$\frac{\omega_{1}\mu_{13}^2+\omega_{2}\mu_{23}^2}{\mu_{13}^2+\mu_{23}^2}$
   &$\frac{\omega_{1}\mu_{23}^2+\omega_{2}\mu_{13}^2}{\mu_{13}^2+\mu_{23}^2}$
   &$\frac{\omega_{2}\mu_{12}^2+\omega_{3}\mu_{13}^2}{\mu_{12}^2+\mu_{13}^2}$
   &$\frac{\omega_{2}\mu_{13}^2+\omega_{3}\mu_{12}^2}{\mu_{12}^2+\mu_{13}^2}$\\
   \hline
   \rule[-4mm]{0mm}{11mm}
   $\tilde{\omega}_{3}$ &$\frac{\omega_{1}\mu_{23}^2+\omega_{3}\mu_{12}^2}{\mu_{12}^{2}+\mu_{23}^2}$
   &$\frac{\omega_{1}\mu_{12}^2+\omega_{3}\mu_{23}^2}{\mu_{12}^{2}+\mu_{23}^2}$
   &{\scriptsize$\omega_{3}$} &{\scriptsize$\omega_{3}$}
   &$\frac{\omega_{2}\mu_{13}^2+\omega_{3}\mu_{12}^2}{\mu_{12}^2+\mu_{13}^2}$
   &$\frac{\omega_{2}\mu_{12}^2+\omega_{3}\mu_{13}^2}{\mu_{12}^2+\mu_{13}^2}$\\
   \hline
   \rule[-4mm]{0mm}{11mm}
   $\tilde{\lambda}_{jk}$ &{\scriptsize$\omega_{31}$}$\frac{\mu_{12}\mu_{23}}{\mu_{12}^{2}+\mu_{23}^2}$
   &-{\scriptsize$\omega_{31}$}$\frac{\mu_{12}\mu_{23}}{\mu_{12}^{2}+\mu_{23}^2}$
   &-{\scriptsize$\omega_{21}$}$\frac{\mu_{13}\mu_{23}}{\mu_{13}^{2}+\mu_{23}^2}$
   &{\scriptsize$\omega_{21}$}$\frac{\mu_{13}\mu_{23}}{\mu_{13}^{2}+\mu_{23}^2}$
   &{\scriptsize$\omega_{32}$}$\frac{\mu_{12}\mu_{13}}{\mu_{12}^2+\mu_{13}^2}$
   &-{\scriptsize$\omega_{32}$}$\frac{\mu_{12}\mu_{13}}{\mu_{12}^2+\mu_{13}^2}$\\
   \hline
   \rule[-4mm]{0mm}{11mm}
   $\tilde{\mu}_{12}$ &{\scriptsize$\sqrt{\mu_{12}^{2}+\mu_{23}^2}$} &$0$ &$0$ &$0$ &{\scriptsize$\sqrt{\mu_{12}^{2}+\mu_{13}^2}$} &$0$\\
   \hline
   \rule[-4mm]{0mm}{11mm}
   $\tilde{\mu}_{13}$ &$0$ &$0$ &$0$ &{\scriptsize$\sqrt{\mu_{13}^{2}+\mu_{23}^2}$} &$0$ &{\scriptsize$\sqrt{\mu_{12}^{2}+\mu_{13}^2}$}\\
   \hline
   \rule[-4mm]{0mm}{11mm}
   $\tilde{\mu}_{23}$ &$0$ &{\scriptsize$\sqrt{\mu_{12}^{2}+\mu_{23}^2}$} &{\scriptsize$\sqrt{\mu_{13}^{2}+\mu_{23}^2}$} &$0$ &$0$ &$0$\\
   \end{tabular}
   \end{center}
   \label{t2a}
\end{table}

\begin{figure}[t]
\begin{center}
\includegraphics[width=0.35\linewidth]{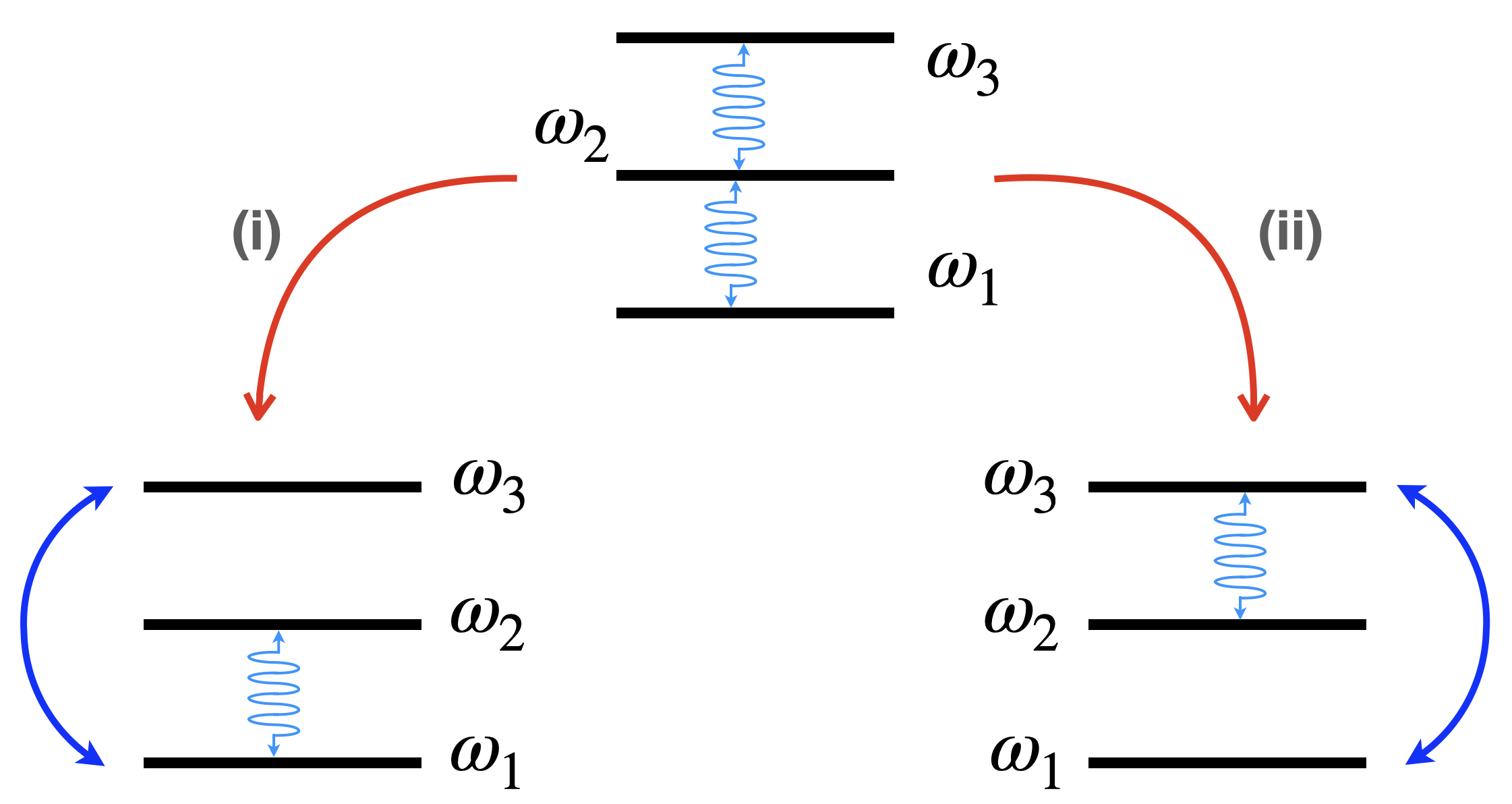}\qquad
\includegraphics[width=0.45\linewidth]{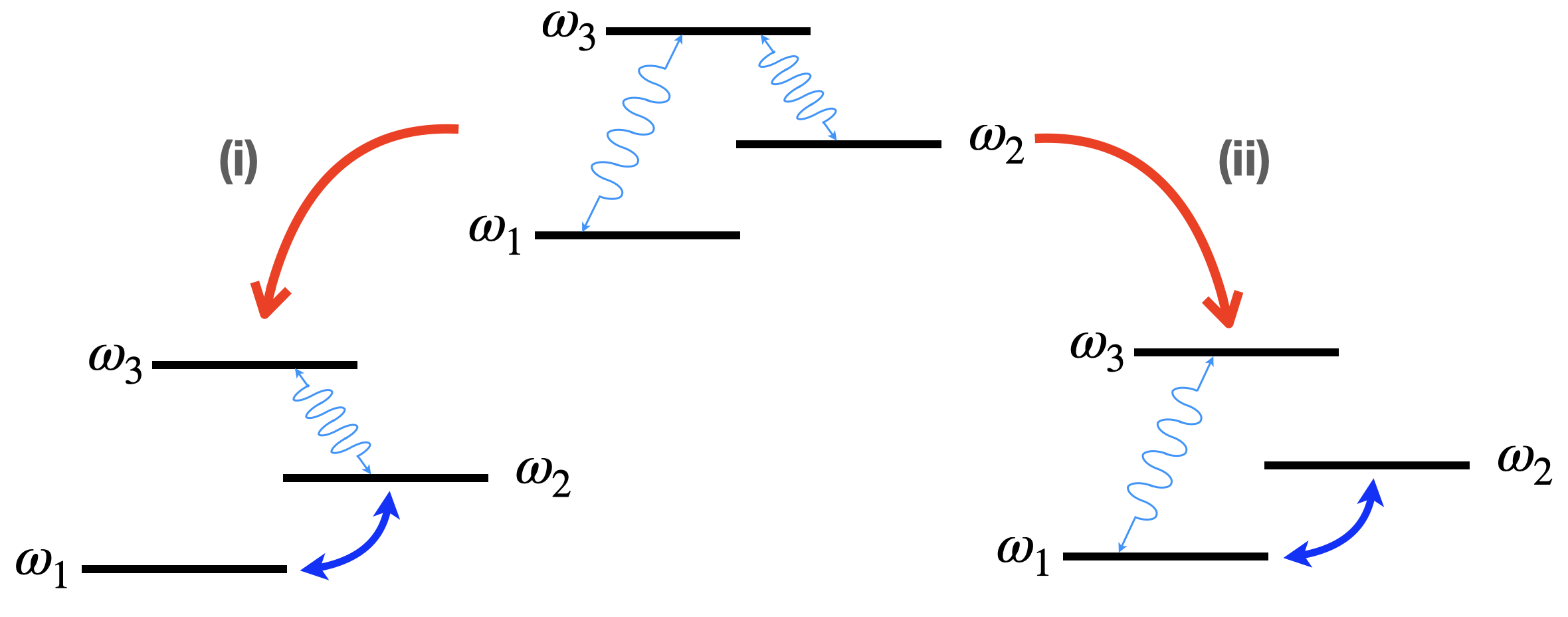}\\[5mm]
\includegraphics[width=0.45\linewidth]{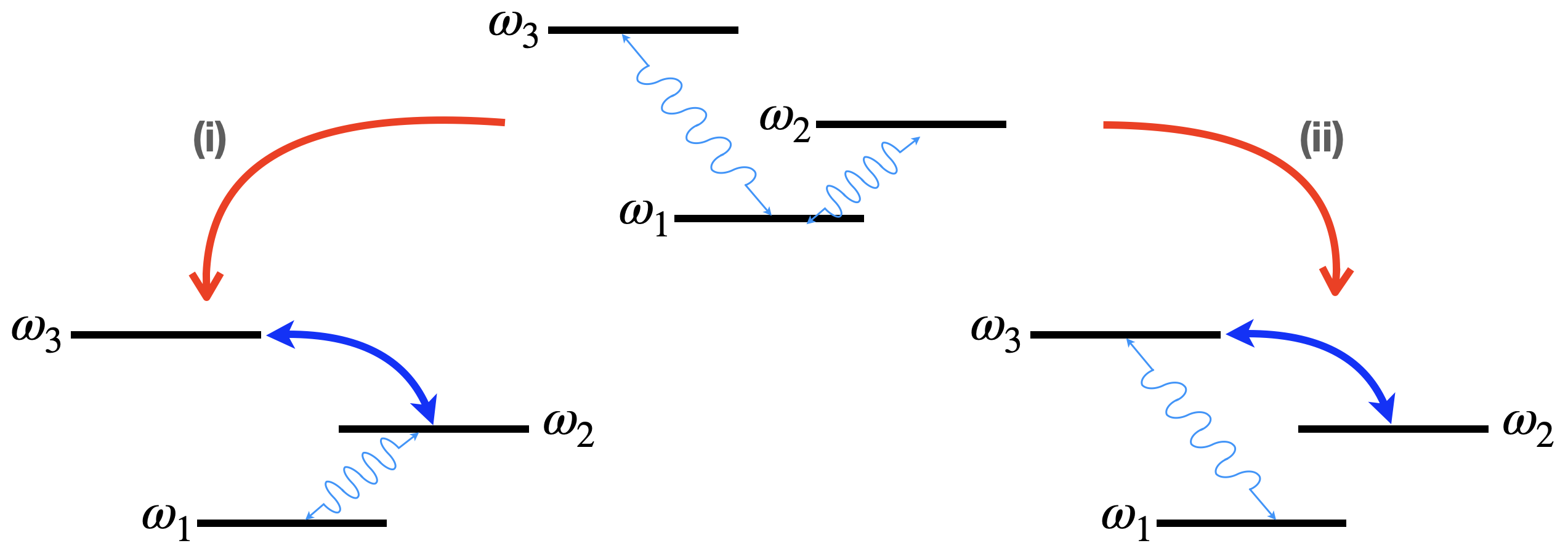}
\caption{Rotations carried out on a $3$-level Dicke Hamiltonian coupled with a one-mode EM-field, in the $\Xi$-, $\Lambda$-, and $V$-configurations, set to eliminate the EM coupling with a pair of atomic level states. For each atomic configuration there are two choices of the parameter $\alpha$, which are shown at down-left and down-right, respectively.}
\label{confXLVfig}
\end{center}
\end{figure}

Figure~\ref{confXLVfig} depicts what the proposed rotation would do; there are two possibilities
for each final result. Although the coupling of the EM field with a pair of atomic levels
states has been eliminated, there is a new matter one-body coupling between the atomic levels without a dipolar transition. The intensity of the one-body matter interaction $\tilde{\lambda}_{jk}$ vanishes in the $\Lambda$- and $V$-configurations when the atomic levels are degenerate (see Table~\ref{t2a}); in this case the $3$-level Dicke model is reduced to an effective $2$-level Dicke model plus an isolated atomic level. This result is independent of the value of the field frequency in the cavity, we denote it as the {\it equal detuning case}. 

This suggests a procedure for storing and retrieving information, or at least for exchanging information between qubits. As an example, lets consider the $\Lambda$-configuration:

Perform a rotation $\op{U}_{12}[\arctan(\mu_{13}/\mu_{23})]$, which isolates level $\omega_1$. This leaves for the ground state a qubit formed by levels $\omega_2 \rightleftharpoons \omega_3$, with a null population expectation value for level $\omega_1$: $\bra\op{A}_{11}\ket=0$. By performing now the rotation $\op{U}_{12}[-\arctan(\mu_{23}/\mu_{13})]\circ\op{U}^{-1}_{12}[\arctan(\mu_{13}/\mu_{23})]$ we switch to a state where level $\omega_2$ is decoupled, the qubit is now formed by $\omega_1 \rightleftharpoons \omega_3$, and the population expectation value $\op{A}_{22}$ of level $\omega_2$ vanishes.

The information has been exchanged between the two qubits and, by symmetry considerations, the fidelity between them is exactly $1$. If one of the qubits were to be read (measured or piped through a communication channel), the stored information in the first one will be retrieved by the second one, through the given rotation applied to the Hamiltonian.

The same process described can be used in a $V$-configuration. However, in the $\Xi$-configuration the population of the ground and excited states do not present this characteristic.

In the example given, the Hamiltonian $\op{H}_\Lambda$ for the original $3$-level Dicke model reduces to an effective Raman transition of a $2$-level Dicke model plus an isolated atomic level for each rotation. In general the basis of the rotated Hamiltonians, formed by a direct product of Fock states are denoted by $|\nu\,;n_j\,,n_k\ket\otimes|n_\ell\ket$, with $\nu$ indicating the photon number; the number of atoms in the atomic levels $\omega_j\,,\omega_k$ and $\omega_\ell$ are related by  $n_a=n_j+n_k$ (number of atoms in the $jk$-subsystem) and $N_a=n_a+n_\ell$ (total number of atoms). Therefore, the Hamiltonian may then be written as $\op{H}_{jk}=\op{h}_{jk} + \tilde{\omega}_\ell \op{A}_{\ell\ell}$, where the subscripts indicate the dipolar transition that remains, and where
\begin{equation}
\op{h}_{jk} = \Omega \op{a}^\dag\op{a} + \tilde{\omega}_j\op{A}_{jj} + \tilde{\omega}_k\op{A}_{kk} - \frac{\mu_{jk}^{\textrm{eff}}}{\sqrt{n_a}} \left(\op{A}_{jk}+\op{A}_{kj}\right)\left(\op{a}^\dag+\op{a}\right)\,,
\end{equation}
stands for a $2$-level effective Dicke Hamiltonian, with $ \mu_{jk}^{\textrm{eff}}:=\sqrt{n_a/N_a}\, \tilde{\mu}_{jk}$.
The corresponding eigenstates take the form
\begin{eqnarray}
|\psi_{jk}(N_a,n_a)\ket\otimes |n_\ell\ket=\sum_{\nu}\sum_{n_s} c_{\nu n_s}|\nu;\,\,n_s,\,n_a-n_s\ket\otimes|n_\ell\ket\,;
\label{nell}
\end{eqnarray}
for the ground state of the system, one has $n_\ell=0$, and the coefficients $c_{\nu\,n_s}$ are obtained by numerical diagonalisation of the matrix Hamiltonian $\op{h}_{jk}$. 

So, restating the procedure mentioned above, and starting from an eigenstate of $\op{H}_\Lambda$, one uses $\op{U}_{12}[\arctan(\mu_{13}/\mu_{23})]$ to store, ($|\psi\ket_\Lambda\to|\psi_{23}\ket\otimes|n_1\ket$), and then $\op{U}_{12}[-\arctan(\mu_{23}/\mu_{13})]$ to retrieve the information ($|\psi_{23}\ket\otimes|n_1\ket \to |\psi_{13}\ket\otimes|n_2\ket$). We may think of the population in the atomic level $\omega_2$ being used as classical bit, in the sense of distinguishing the states with zero occupation and non-zero occupation in the atomic level $\omega_2$.

{\it Remark:} In a dynamic scenario, by initially preparing the state as \[\vert\psi(t=0)\rangle = |\nu;\,n_2=0,\,n_3=1\rangle\otimes|n_1=0\rangle\,,\] performing the first rotation $\op{U}_{12}[\arctan(\mu_{13}/\mu_{23})]$ will lead to quantum Rabi oscillations in the qubit formed by levels $\omega_2 \rightleftharpoons \omega_3$, with $\bra\op{A}_{11}\ket=0$; by performing the second rotation $\op{U}_{12}[-\arctan(\mu_{23}/\mu_{13})]\circ\op{U}^{-1}_{12}[\arctan(\mu_{13}/\mu_{23})]$ we switch to Rabi oscillations in the qubit $\omega_1 \rightleftharpoons \omega_3$, with now $\op{A}_{22}$ vanishing. Once again, if one of the qubits were to be read, the {\it stored} information in the first one will be {\it retrieved} by the second one.

\section{Numerical Results}

As was pointed out above, the special rotations which permit to reduce the $3$-level Dicke model to an effective $2$-level Dicke model plus an isolated atomic level, may be used as a tool for storing and retrieving qubits. 

In order to study the robustness of this result, it is of interest to consider the case where the atomic levels are not degenerate (different detuning value), i.e., to study as a function of the control parameters  the effect due to the one-body matter interaction. The strength $\tilde{\lambda}_{jk}$ couples the states which are not affected by  the dipolar coupling with the EM field, one of them associated to the state uncoupled by the rotation.

\subsection{Level population}

\begin{figure}
	\includegraphics[scale=0.23]{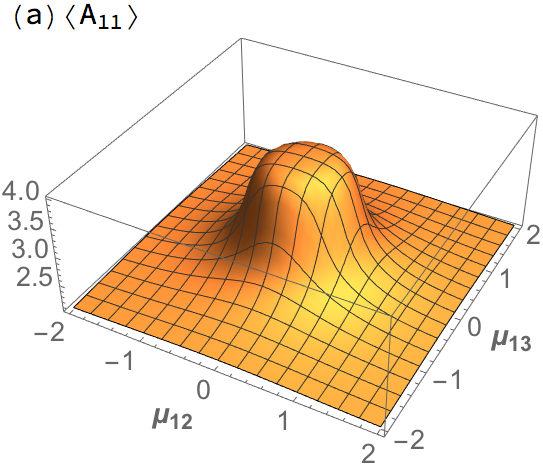}\quad
	\includegraphics[scale=0.23]{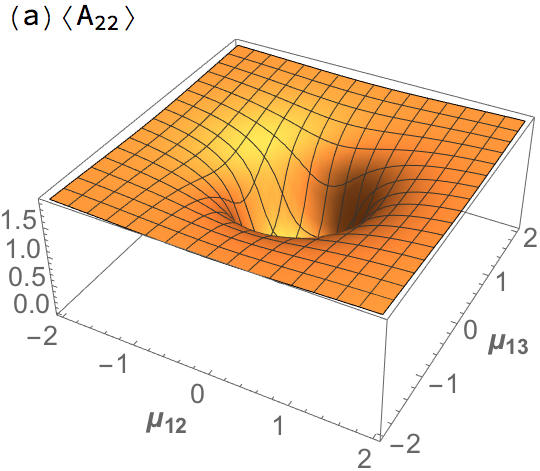}\quad
	\includegraphics[scale=0.26]{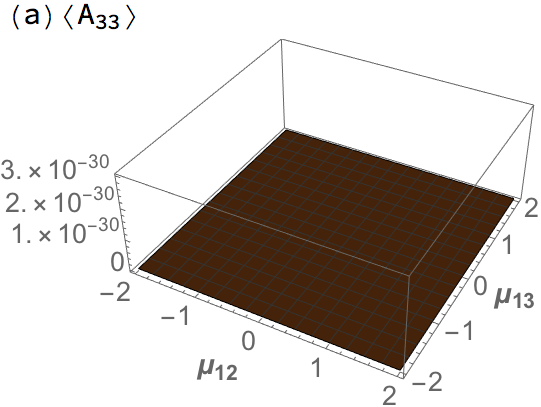}\\
	\includegraphics[scale=0.23]{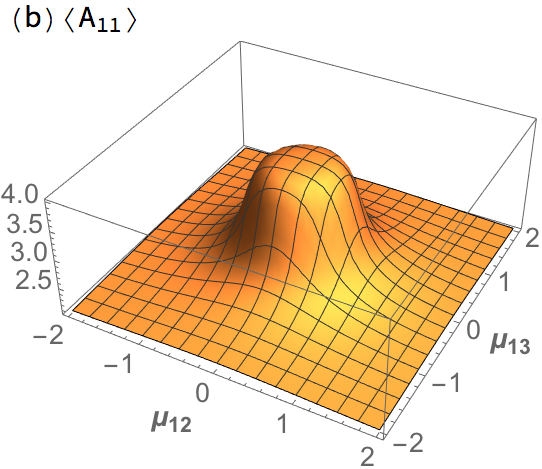}\quad
	\includegraphics[scale=0.26]{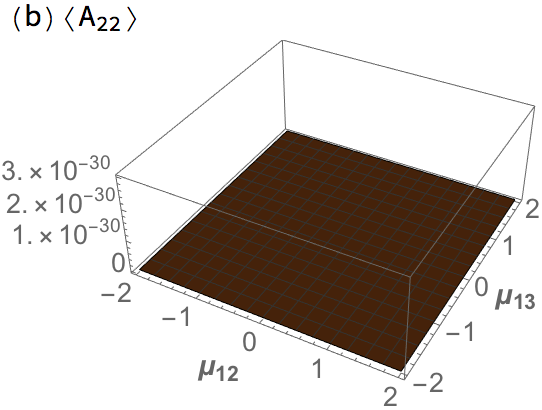}\quad
	\includegraphics[scale=0.23]{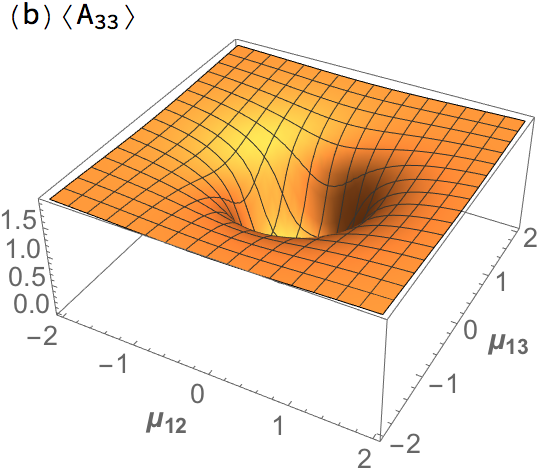}
	\caption{ Expectation values for the population of the atomic levels $\omega_1$, $\omega_2$, and $\omega_3$: (a) Storing stage,  after the $\mathfrak{su}(3)$ rotation~(\ref{su3rotation}) with $\tan(\alpha)=\mu_{13}/\mu_{12}$. (b) Retrieving stage, after the $\mathfrak{su}(3)$ rotation~(\ref{su3rotation}) with $\tan(\alpha)=-\mu_{13}/\mu_{12}$. Paramenters are: $\omega_1=\omega_2=0$, $\omega_3=1$, $\Omega=1$ and $N_a=4$.
	}
	\label{FigurasConPoblacionCero}
\end{figure}

It is interesting to calculate the expectation value of the population for each atomic level under the special rotations shown in Table~\ref{t2a}. We find that, when we have equally detuned energy levels in the $\Lambda$- and $V$-configurations, the population of the isolated level is exactly naught, as shown representatively for the $V$-configuration in Fig.~\ref{FigurasConPoblacionCero}. Notice that the expectation value of the population in the lowest atomic level $\bra \op{A}_{11}\ket$ remains unaltered under these rotations; this is due to the fact that the rotation $\op{U}_{23}$ only affects the population of the atomic levels $\omega_2$ and $\omega_3$.

%

\begin{figure}[t]
\includegraphics[width=0.3\linewidth]{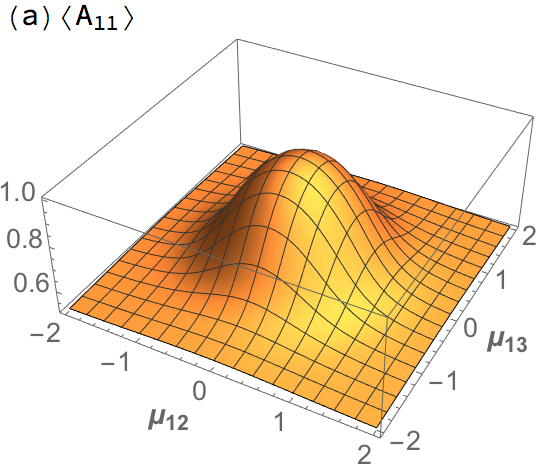}\quad
\includegraphics[width=0.3\linewidth]{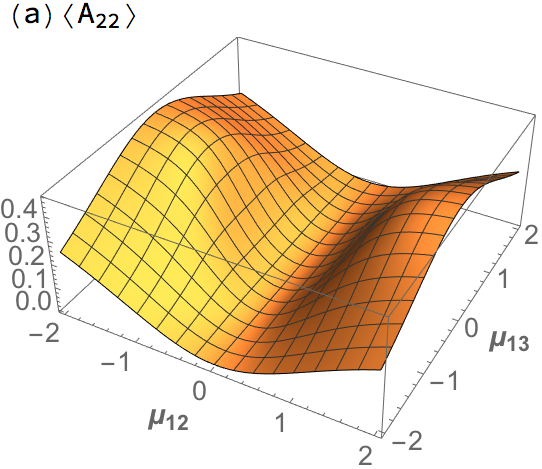}\quad
\includegraphics[width=0.3\linewidth]{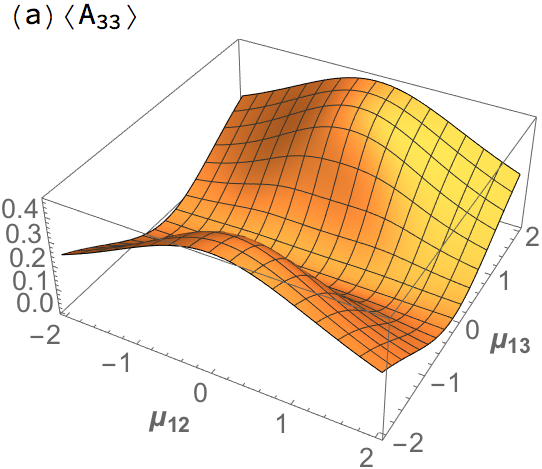}\\
\includegraphics[width=0.3\linewidth]{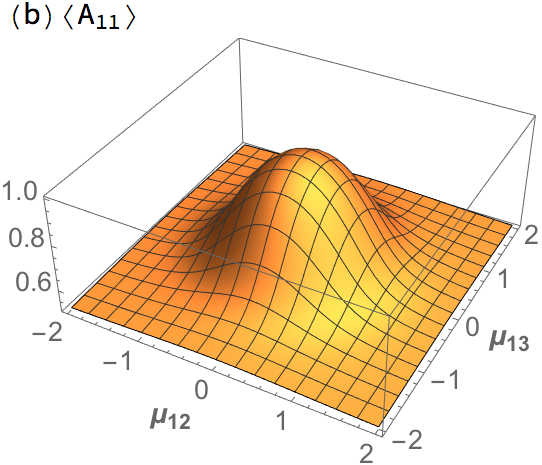}\quad
\includegraphics[width=0.3\linewidth]{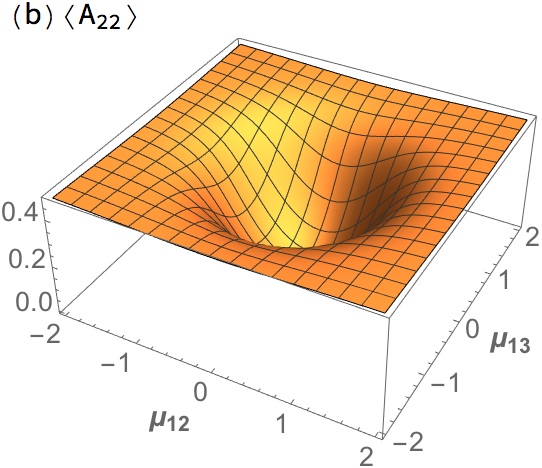}\quad
\includegraphics[width=0.3\linewidth]{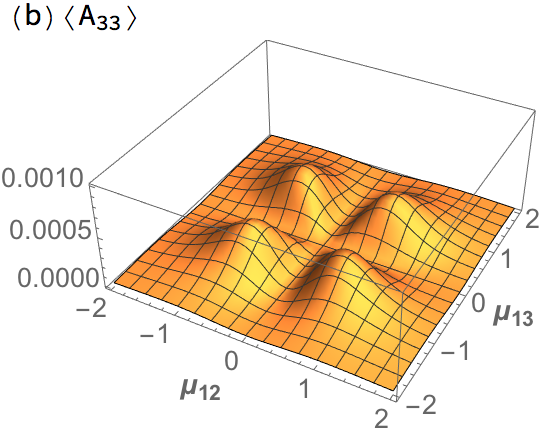}\\
\includegraphics[width=0.3\linewidth]{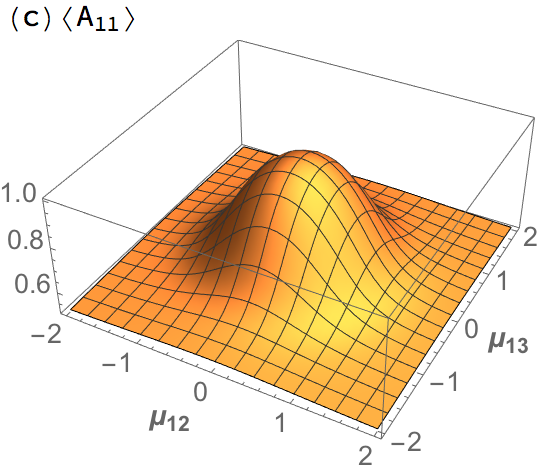}\quad
\includegraphics[width=0.3\linewidth]{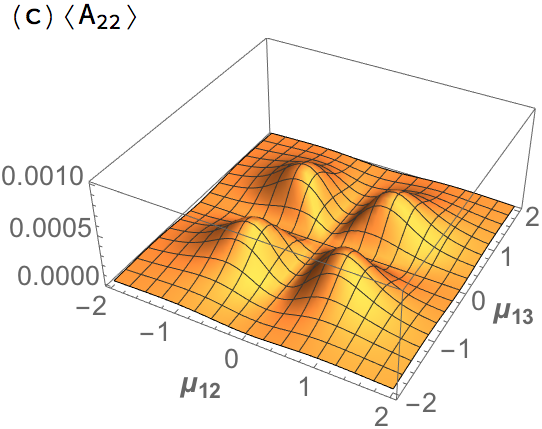}\quad
\includegraphics[width=0.3\linewidth]{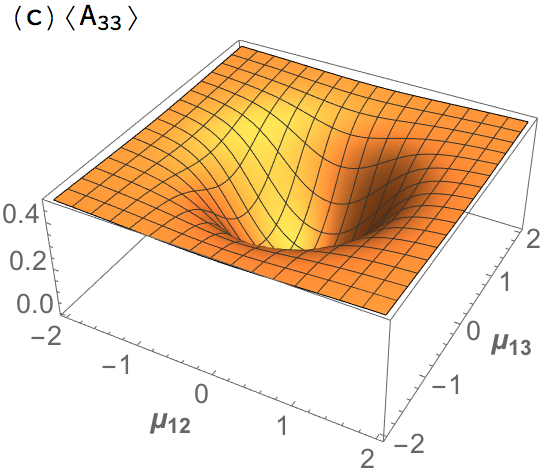}
\caption{Expectation values of the atomic populations $\langle A_{11}\rangle$, $\langle A_{22}\rangle$, and $\langle A_{33}\rangle$, as functions of the control parameters, for a single atom in the $V$-configuration, with parameters $\omega_1=0$, $\omega_2=0.8$,  $\omega_3=1$ and $\Omega=1$.
(a) The quantum ground state of the Hamiltonian without rotation $\op{H}_V$.
(b) The ground state of the Hamiltonian $\op{H}_{12}$ Eq.~(\ref{eq.H12}). (c) The ground state of the Hamiltonian $\op{H}_{13}$ Eq.~(\ref{eq.H13}). Notice that the population of the uncoupled levels are very close to zero.}
\label{f.evAjj}
\end{figure}

Even in the situation when the two levels are not in equal detuning, say off by as much as $20\%$, the expectation value of the population of the isolated level is extremely small and can reach up to $4\times 10^{-4}$, as figure~\ref{f.evAjj} will show below. This makes the fidelity between states less than one, but as we approach resonance, the fact that the retrieved photon state is identical to the stored one makes the fidelity between the two approach to~$1$.

In both the $\Lambda$- and $V$-configurations under degenerate atomic level conditions (equal detuning values), the one-body matter interaction term vanishes. When not in equal detuning, the one-body matter interaction term has an effect over the isolated atomic level. We found by numerical diagonalisation of the Hamiltonians that both configurations have similar results. As a numerical example consider the atomic $V$-configuration, the Hamiltonian of which without rotations $\op{H}_V$ is given by Eq.~(1) with $\mu_{23}=0$, indicating that the transition $\omega_2\rightleftharpoons\omega_3$ is prohibited. Under the rotations considered, the Hamiltonian takes the form
\begin{eqnarray}
\op{H}_{12} &=& \op{h}_{12} + \tilde{\omega}_3\op{A}_{33} + \tilde{\lambda}_{23}\left(\op{A}_{23}+\op{A}_{32}\right)\,,\label{eq.H12}\\[2mm]
\op{H}_{13} &=& \op{h}_{13} + \tilde{\omega}_2\op{A}_{22} + \tilde{\lambda}_{23}\left(\op{A}_{23}+\op{A}_{32}\right)\,.\label{eq.H13}
\end{eqnarray}
%

%

In figure~\ref{f.evAjj} the atomic populations of the ground state as function of the control parameters $\mu_{12}$ and $\mu_{13}$ are shown for a single atom $N_a=1$. As expected, the population of the atomic level $\omega_1$ is close to unity in the normal region (close to the origin), while for the collective region this decreases to $\bra \op{A}_{11}\ket \approx 0.5$, indicating that in this region the other atomic levels contribute an important role to the ground state. The figure also shows the corresponding atomic populations for the transformed Hamiltonians Eqs.~(\ref{eq.H12}) and (\ref{eq.H13}), respectively; while the atomic population of level $\omega_1$ remains unaltered, those of the higher atomic levels change drastically; in particular, we note that the higher atomic level populations exchange their role under these rotations.

It is of interest to note that the expectation value $\bra\op{A}_{\ell\ell}\ket$ of the atomic population of the atomic level that decouples  after the rotation remains close to zero; in fact, while the numerical calculation considers a $20\%$ in the atomic level detuning, we obtain a deviation of less than $0.05\%$ in the atomic population of this decoupled state, showing clearly that this property of the rotations is robust.

\subsection{Phase Diagram}

The separatrix of a system is characterised by abrupt changes in the ground state of the system when a control parameter is varied (cf. e.g. \cite{sachdev11}). In order to detect these loci we use the fidelity between neighbouring ground states~\cite{zanardi06}
\begin{equation}
F(\mu,\mu+\delta\mu):= | \langle \psi(\mu)|\psi(\mu+\delta\mu)\rangle|^2\,,
\end{equation}
so that points where a minimum of the fidelity occurs characterise abrupt changes in the state of the system. 

The same procedure can be applied to excited quantum states  to determine the quantum phase transitions of the excited states, as has been studied in other quantum scenarios, such as molecular linear-to-bent transitions~\cite{perez-bernal08}. For a review of excited-state quantum phase transitions, see e.g.~\cite{cejnar21}.

Since in our system we have two control parameters, we take a pencil of straight lines based at the origin and find the points where a minimum occurs along these lines. Clearly, the precision in the location of the points will depend on the step $\delta\mu>0$ taken.

We can show, firstly, that for all three atomic configurations, the separatrix for the two particular rotations that we are applying to the generalised Dicke model does not change with respect to that of the original Hamiltonian. In~\ref{AppB} we calculate the fidelity between neighbouring states as
   \begin{eqnarray} \label{fidela}
      &&F_{\rm rot}(\mu,\mu+\delta\mu)\approx |\langle\psi(\mu+\delta\mu)|\psi(\mu)\rangle|^{2}\\
      &&\quad+(\delta\mu)^{2}\,\left(\frac{d\alpha}{d\mu}\right)^{2}\,\Big[\langle\psi(\mu+\delta\mu)|\psi(\mu)\rangle
      \langle\psi(\mu+\delta\mu)|\op{K}^{2}|\psi(\mu)\rangle
      +|\langle\psi(\mu+\delta\mu)|\op{K}|\psi(\mu)\rangle|^{2}\Big]\ ,\nonumber
   \end{eqnarray}
where the operator $\op{K}$ stands for $\op{K}_{jk}$ Eq.~(\ref{eq.Kjk}): $\op{K}_{31}=\op{A}_{31}-\op{A}_{13}$ for the $\Xi$-configuration, $\op{K}_{12}=\op{A}_{12}-\op{A}_{21}$ for the $\Lambda$-configuration, and $\op{K}_{32}=\op{A}_{32}-\op{A}_{23}$ for the $V$-configuration. In fact, it is only necessary to calculate $d\alpha/d\mu$; the results are shown in Table~\ref{t3}.

The phase diagram is invariant under ${\rm SU}(2)$ rotations, even when these depend on the control parameters (cf. \ref{AppB}).  We have found that, although the value of the fidelity between neigbouring states is not the same, the stability and equilibrium properties are the same for the original and the rotated Hamiltonians (cf. Figs.~\ref{f.sepX}-\ref{f.sepL} below).

\begin{table}[h]
   \caption{Values of derivatives which enter in Eq.~(\ref{fidela}), for each configuration, when calculated maintaining one
   $\mu_{jk}$ fixed while varying the other. }
   \begin{center}
   \begin{tabular}{c|c|c||c|c|c||c|c|c}
   &\multicolumn{2}{|c||}{$\Xi$}&&\multicolumn{2}{c||}{$\Lambda$}&&\multicolumn{2}{|c}{$V$}\\
   \hline\\[-5mm]
   \hline
   \rule[-4mm]{0mm}{11mm}
   $\tan(\alpha)$ &$\frac{\mu_{23}}{\mu_{12}}$ &-$\frac{\mu_{12}}{\mu_{23}}$ &&$\frac{\mu_{13}}{\mu_{23}}$ 
   &-$\frac{\mu_{23}}{\mu_{13}}$ &&$\frac{\mu_{13}}{\mu_{12}}$ &-$\frac{\mu_{12}}{\mu_{13}}$\\
   \hline
   \rule[-4mm]{0mm}{11mm}
   $\frac{d\alpha}{d\mu_{12}}$ &\multicolumn{2}{|c||}{-$\frac{\mu_{23}}{\mu_{12}^{2}+\mu_{23}^{2}}$}
   &$\frac{d\alpha}{d\mu_{23}}$ &\multicolumn{2}{|c||}{-$\frac{\mu_{13}}{\mu_{13}^{2}+\mu_{23}^{2}}$}
   &$\frac{d\alpha}{d\mu_{13}}$ &\multicolumn{2}{|c}{$\frac{\mu_{12}}{\mu_{12}^{2}+\mu_{13}^{2}}$}\\
   \hline
   \rule[-4mm]{0mm}{11mm}
   $\frac{d\alpha}{d\mu_{23}}$ &\multicolumn{2}{|c||}{$\frac{\mu_{12}}{\mu_{12}^{2}+\mu_{23}^{2}}$}
   &$\frac{d\alpha}{d\mu_{13}}$ &\multicolumn{2}{|c||}{$\frac{\mu_{23}}{\mu_{13}^{2}+\mu_{23}^{2}}$}
   &$\frac{d\alpha}{d\mu_{12}}$ &\multicolumn{2}{|c}{-$\frac{\mu_{13}}{\mu_{12}^{2}+\mu_{13}^{2}}$}
   \end{tabular}
   \end{center}
   \label{t3}
\end{table}

The calculation of the fidelity between neigbouring states along lines maintening one coupling strength constant (as in Table~\ref{t3}) does not always detect all the minima; however, along linear and/or quadratic functions that pass through the origin all minima are determined.

Fig.~\ref{f.sepX} shows the phase diagram, given by the minimum of the fidelity between neighbouring states, in the $\Xi$-configuration, for the ground-state of Hamiltonian~(\ref{ham3p}) in a resonant case with $\omega_{1}=0$, $\omega_{2}=1$, $\omega_{3}=2$ and $\Omega=1$ Fig.~\ref{f.sepX}(a), and a non-resonant case in which $\omega_{2}=1.5$ Fig.~\ref{f.sepX}(b), for different number of atoms $N_{a}=1,\,2,\,3$ and $4$.  In both cases, these are compared with the separatrix obtained from the variational solution, using coherent states as test functions~\cite{cordero13a,cordero13b}, which has the analytical expression
\begin{equation}\label{eq.sepX}
\Omega \omega_{21} = 4 \mu_{12}^2 + \left[2|\mu_{23}|-\sqrt{\Omega \omega_{31}}\right]^2 \Theta(2|\mu_{23}|-\sqrt{\Omega \omega_{31}})\,.
\end{equation}
Here $\Theta(x)$ stands for the Heaviside theta function, and this variational separatrix remains fixed for any number of atoms $N_a$ (continuous, magenta line). The factor of $2$ that multiplies the dipolar strength $\mu_{jk}$ arises due to the fact that we have not considered the RWA approximation. Notice that, as a function of the number of atoms $N_a$, the quantum separatrix tends to that obtained from the variational solution as $N_a$ increases. Also, the variational separatrix lies very near the quantum calculation around $\mu_{23}$-axis, this is due to the fact that in the normal region the quantum ground state is characterised by states which are close to both, the vacuum field contribution and the matter ground state, so in order to transit from the normal to the collective region in that direction requires large values of the dipolar intensity $\mu_{23}$, a regime where the variational solution provides the best approach to the quantum ground state energy. In addition, the effect on the separatrix when we consider a different value of the detuning is to augment the normal region (compare the region limited by the variational separatrix (continuous, magenta line) in figures~\ref{f.sepX}(a) and \ref{f.sepX}(b)). The separatrix for this atomic configuration has the same shape independently of the detuning values Eq.~(\ref{eq.detuning}). 

%
\begin{figure}
\includegraphics[width=0.45\linewidth]{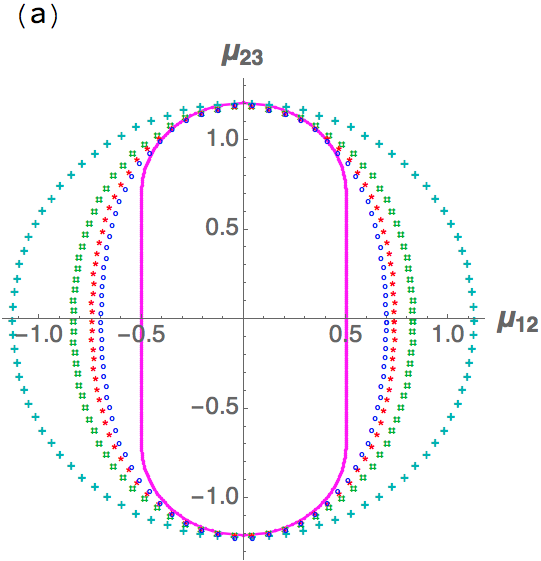}\quad
\includegraphics[width=0.45\linewidth]{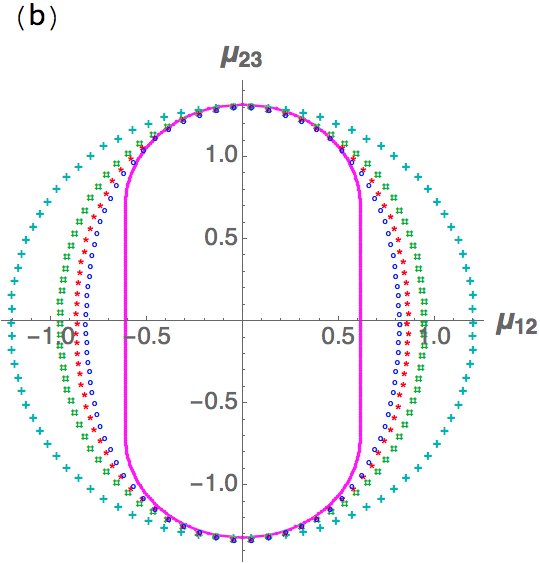}
\caption{Separatrix for atoms in the $\Xi$-configuration as a function of the number of atoms $N_a=1,\,2,\,3$ and $4$ (dotted shapes) in comparison with the corresponding variational solution (continuous line). The parameters are: $\omega_1=0,\,\omega_3=2$ for the atomic levels and $\Omega=1$ for the field frequency, and the cases considered are (a) equal detuning $\omega_2=1$ and (b) different detuning with $\omega_2=1.5$. As the number of atoms increases the quantum separatrix tends to the variational one.}
\label{f.sepX}
\end{figure}

Figure~\ref{f.sepV} shows the separatrices corresponding to the atomic $V$-configuration for $N_a=1,\,2,\,3\,$ and $4$ atoms, with zero detuning Fig.~\ref{f.sepV}(a), and non-zero detuning Fig.~\ref{f.sepV}(b). The variational separatrix obeys in this case the equation
\begin{equation}\label{eq.sepV}
\frac{ 4\mu_{12}^2}{\Omega\omega_{21}} + \frac{ 4\mu_{13}^2}{\Omega\omega_{31}}=1\,,
\end{equation}
so for an equal detuning we have a circumference, while in other cases the separatrix has an elliptic shape, which changes for different detunings.

\begin{figure}
\includegraphics[width=0.45\linewidth]{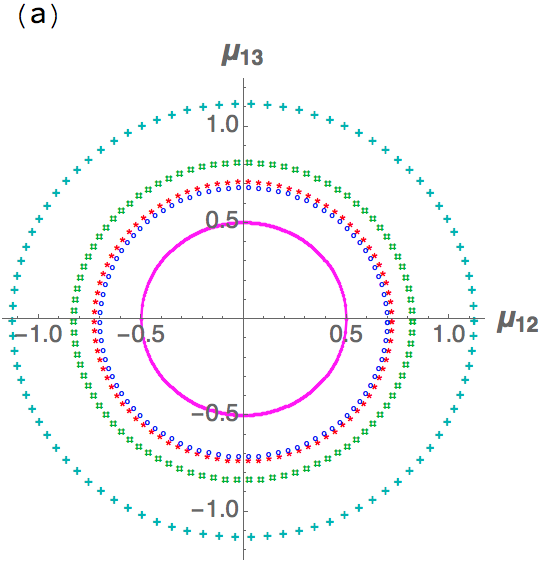}\quad
\includegraphics[width=0.45\linewidth]{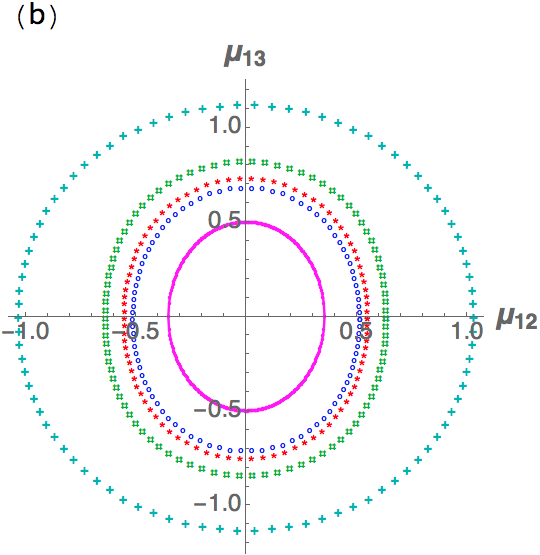}
\caption{Separatrix for atoms in the atomic $V$-configuration as a function of the number of atoms $N_a=1,\,2,\,3$ and $4$ (dotted shapes) these are compared with the corresponding variational solution (continuous line). The parameters are: $\omega_1=0,\,\omega_3=1$ for the atomic levels and $\Omega=1$ for the field frequency, for the cases (a) equal detuning $\omega_2=1$ and (b) different detuning with $\omega_2=0.5$. As the number of atoms increases the quantum separatrix tends to the variational one.}
\label{f.sepV}
\end{figure}

In a similar fashion, the separatrices for atoms in the $\Lambda$-configuration are shown in figure~\ref{f.sepL}, with equal detuning Fig.~\ref{f.sepL}(a) and different detuning Fig.~\ref{f.sepL}(b). For this atomic configuration one finds that the variational separatrix (continuous line) is given by
\begin{equation}\label{eq.sepL}
\Omega \omega_{31} = 4\mu_{13}^2 + \left[2|\mu_{23}|-\sqrt{\Omega \omega_{21}}\right]^2 \Theta(2|\mu_{23}|-\sqrt{\Omega \omega_{21}})\,.
\end{equation}
Clearly, for the case of equal detuning the shape is that of a circumference, similar to the atomic $V$-configuration, since equal detuning is obtained when $\omega_{21}=0$ (cf. Eq.~(\ref{eq.detuning})). For a different detuning one has $\omega_{21}\neq0$, and the variational separatrix imitates the separatrix shape of the atoms in the $\Xi$-configuration. In this case, as $N_a$ increases, the quantum separatrix tends to the variational result along the $\mu_{23}$-axis {\it from} the normal {\it to} the collective region. 

%
\begin{figure}
\includegraphics[width=0.45\linewidth]{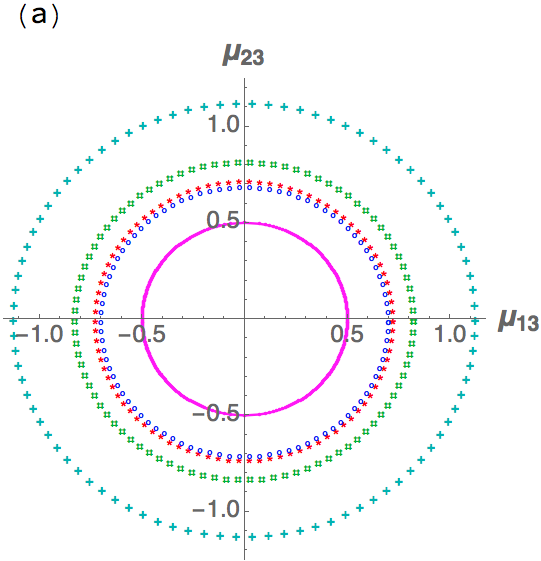}\quad
\includegraphics[width=0.45\linewidth]{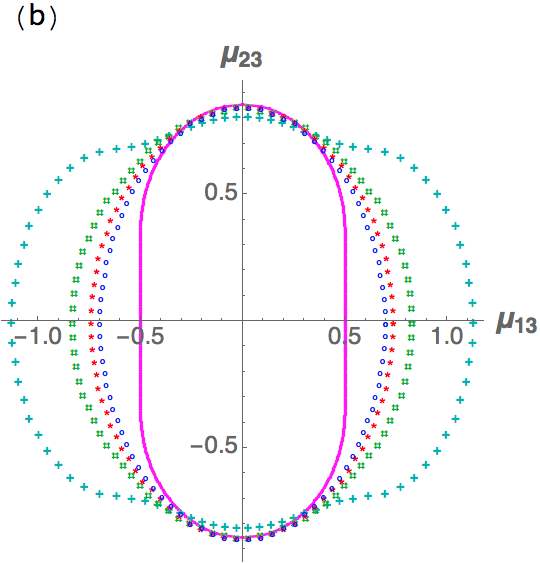}
\caption{Separatrix for atoms in the $\Lambda$-configuration as function of the number of atoms $N_a=1,\,2,\,3$ and $4$ (dotted shaps) in comparison with the corresponding variational solution (full line). The parameters of system are: $\omega_1=0,\,\omega_3=1$ for the atomic levels and $\Omega=1$ field frequency, for (a) equal detuning $\omega_2=0$ and (b) different detuning with $\omega_2=0.5$. As the number of atoms increases the quantum separatrix goes to the variational one.}\label{f.sepL}
\end{figure}

\section{Conclusions}

We have proposed a methodology that could possibly be used for retrieving stored quantum information via a simple rotation applied to the Hamiltonian, in atomic systems interacting dipolarly with a one-mode radiation field. As a numerical example, and for robustness of our result,
we considered the atomic $V$-configuration when not in equal detuning, in which case the population $\langle A_{33} \rangle$  or $\langle A_{22} \rangle$ are not exactly naught; for a detuning of $20\%$, for instance, the population will reach up to $0.05\%$ (cf. Fig.~\ref{f.evAjj}), which makes the fidelity between states less than one. But as we approach resonance, from symmetry arguments, the fact that the ``retrieved'' photon state is identical to the ``stored'' one makes the fidelity between the two approach $1$. The same construct can be carried out with atoms (or artificial atoms) in the atomic $\Lambda$-configuration.

Additionally, this method allows for the obtention of the full energy spectrum of the system, and write it in bands associated to the number of particles $n_\ell = 0,1,\ldots,N_a$ of the isolated Fock state, as shown in eq.~(\ref{nell}). The parity symmetry allows us to write the Hilbert space as $\mathcal{H} = \mathcal{H}_e \oplus \mathcal{H}_o$, with $\mathcal{H}_e$ formed by the states of an even number of total excitations $M$ (eigenvalue of the operator $\op{M}_V$ or $\op{M}_{\Lambda}$), and $\mathcal{H}_o$ by states of an odd number of excitations. The transition operators between bands are of the form $\op{A}_{k\ell}$ and $\op{A}_{\ell k}$, with the restriction that $N_a=n_{\ell}+n_k+n_j$ remains constant.

Furthermore, given a fixed number of total excitations $M$, the method considerable reduces the dimension of the space under study, by {\it freezing} one of the three atomic levels and reducing it to a two-level Dicke system. In the $V$-configuration case, this two-level system will have both, even and odd parities depending on the parity of the frozen level; on the other hand, in the $\Lambda$-configuration the parity of the two level system is independent of the parity of the frozen level.

\subsection*{Acknowledgments}
This work was partially supported by DGAPA-UNAM under project IN100323.

\appendix
\section{Rotation of the 3-Level Dicke Hamiltonian with one EM-mode for arbitrary $\alpha$}
\label{AppA}

From Table~\ref{t1} one may consider the different rotations using the ${\rm SU}(2)$ relevant groups for each $3$-level atomic configuration. In particular, the operator of the form
\begin{equation}
\op{K}_{jk} := \op{A}_{jk}-\op{A}_{kj}\,; \qquad \op{K}_{jk}^\dag =-\op{K}_{jk}\,,
\end{equation}
calls our attention, from which one has the rotation of the form
\begin{equation}
\op{U}_{jk}(\alpha):= e^{-\alpha \op{K}_{jk}}\,,\qquad \alpha^*=\alpha\,.
\end{equation}

We denote  $\tilde{\op{A}}_{\ell m}(j,k) := \op{U}_{jk} \op{A}_{\ell m} \op{U}_{jk}^\dag$ the transformed operators, for fixed $j$ and $k$ with $j\neq k$. Clearly, when  $\ell\neq j,\,\ell\neq k$ y $m\neq j,\,m\neq k$ one has $\tilde{\op{A}}_{\ell m}(j,k)  =  \op{A}_{\ell m}$. In order to find the transformed operators  in terms of the original ones, with at least one coincidental index with the transformation, we use a well-known method, which consists on finding a differential equation for each operator~\cite{miller72,cohen20}. 

The first derivative yields a coupled system of equations
\begin{eqnarray*}
\frac{\partial}{\partial\alpha}\tilde{\op{A}}_{\ell m}(j,k) 
&=&  -\left(\delta_{\ell k}\tilde{\op{A}}_{jm}(j,k) -\delta_{jm}\tilde{\op{A}}_{\ell k} (j,k) -  \delta_{\ell j}\tilde{\op{A}}_{km}(j,k) +\delta_{km}\tilde{\op{A}}_{\ell j}(j,k) \right)\,,
\end{eqnarray*}
and taking the second derivative one finds a second order differential equation for each genertor of the $\mathfrak{su}(3)$ algebra. The transformed operators (for fixed $j$ and $k$, $j\neq k$) take the form
\begin{eqnarray*}
\tilde{\op{A}}_{jj}(j,k)  &=& \cos^2(\alpha) \op{A}_{jj} + \sin^2(\alpha) \op{A}_{kk} + \cos(\alpha)\sin(\alpha)\left(\op{A}_{jk}+\op{A}_{kj}\right)\,,\\[2mm]
\tilde{\op{A}}_{kk}(j,k)  &=& \cos^2(\alpha) \op{A}_{kk} + \sin^2(\alpha) \op{A}_{jj} - \cos(\alpha)\sin(\alpha)\left(\op{A}_{jk}+\op{A}_{kj}\right)\,,\\[2mm]
\tilde{\op{A}}_{jk}(j,k)  &=& \cos^2(\alpha) \op{A}_{jk} - \sin^2(\alpha) \op{A}_{kj} + \cos(\alpha)\sin(\alpha)\left(\op{A}_{kk}-\op{A}_{jj}\right)\,,\\[2mm]
\tilde{\op{A}}_{kj}(j,k)  &=& \cos^2(\alpha) \op{A}_{kj} - \sin^2(\alpha) \op{A}_{jk} + \cos(\alpha)\sin(\alpha)\left(\op{A}_{kk}-\op{A}_{jj}\right) \,,\\[2mm]
\tilde{\op{A}}_{jm}(j,k)  &=& \cos(\alpha) \op{A}_{jm} + \sin(\alpha)\op{A}_{km}\,,\\[2mm]
\tilde{\op{A}}_{km}(j,k) &=& \cos(\alpha) \op{A}_{km} - \sin(\alpha)\op{A}_{jm}\,,\\[2mm]
\tilde{\op{A}}_{\ell j}(j,k)  &=& \cos(\alpha) \op{A}_{\ell j} + \sin(\alpha)\op{A}_{\ell k}\,,\\[2mm]
\tilde{\op{A}}_{\ell k}(j,k)  &=& \cos(\alpha) \op{A}_{\ell k} - \sin(\alpha)\op{A}_{\ell j}\,.
\end{eqnarray*}
These transformations are valid for an $n$-dimensional algebra. 

For each atomic configuration, this special rotation is defined by the forbidden atomic transition, that is, for the $\Xi$-configuration one has $\op{U}_{31}(\alpha)$, for the $\Lambda$-configuration $\op{U}_{12}(\alpha)$ and for the $V$-configuration the rotation is  $\op{U}_{32}(\alpha)$. The advantage of this kind of rotation is the fact that, for an appropriate value of $\alpha$, one may eliminate one term of the matter-field interaction in the $3$-level Dicke model with a single electromagnetic field.  So, for atoms in the $\Xi$-configuration one has 
   \begin{eqnarray*}
      \op{H}_{\Xi}^{\prime}(\alpha)&=&\Omega\,\op{a}^{\dagger}\op{a}+\left(\omega_{1}\,\cos^{2}(\alpha)+\omega_{3}
      \sin^{2}(\alpha)\right)\,\op{A}_{11}+\omega_{2}\,\op{A}_{22}\nonumber\\
      &+&\left(\omega_{1}\,\sin^{2}(\alpha)+\omega_{3}\,\cos^{2}(\alpha)\right)\,\op{A}_{33}
      +\frac{1}{2}\,\omega_{31}\,\sin(2\,\alpha)
      \left(\op{A}_{13}+\op{A}_{31}\right)\nonumber\\
      &-&\frac{1}{\sqrt{N_{a}}}\left(\op{a}^{\dagger}+\op{a}\right)\Big\{
      \left(\cos(\alpha)\,\mu_{12}+\sin(\alpha)\,\mu_{23}\right)
      \left(\op{A}_{12}+\op{A}_{21}\right)\nonumber\\
      &&\phantom{-\frac{1}{\sqrt{N_{a}}}(\op{a}^{\dagger})}
      +\left(-\sin(\alpha)\,\mu_{12}+\cos(\alpha)\,\mu_{23}\right)
      \left(\op{A}_{23}+\op{A}_{32}\right)\Big\}\ .
      \label{hamXrot}
   \end{eqnarray*}
and taking the value of $\alpha$ such that
\begin{eqnarray*}
\cos(\alpha)\mu_{12} + \sin(\alpha)\mu_{23} =0\,, \quad \textrm{or}\quad 
-\sin(\alpha)\mu_{12} + \cos(\alpha)\mu_{23} =0\,,
\end{eqnarray*}
eliminates the dipolar transition $\omega_1\rightleftharpoons \omega_2$ or  $\omega_2\rightleftharpoons \omega_3$, respectively. For the particular values $\alpha=0$ and $\alpha=\pi/2$ the one-body matter interaction vanishes.

For the $\Lambda$-configuration the Hamiltonian is transformed to
\begin{eqnarray*}
\op{H}'_\Lambda &=& \Omega\op{a}^\dag\op{a} + \left[\omega_1\cos^2(\alpha) + \omega_2\sin^2(\alpha)\right]\op{A}_{11} + \left[\omega_1\sin^2(\alpha) + \omega_2\cos^2(\alpha)\right]\op{A}_{22} + \omega_3\op{A}_{33} \\[2mm]
&-& \frac{1}{2}\omega_{21} \sin(2\alpha)\left(\op{A}_{12}+\op{A}_{21}\right) - \frac{1}{\sqrt{N_a}}\left(\op{a}^\dag+\op{a}\right) \\[2mm]
&\times& \Big\{ \left[\cos(\alpha)\mu_{13} -\sin(\alpha)\mu_{23}\right]\left(\op{A}_{13}+\op{A}_{31}\right) + \left[\sin(\alpha)\mu_{13} +\cos(\alpha)\mu_{23}\right]\left(\op{A}_{23}+\op{A}_{32}\right)\Big\}\,,
\end{eqnarray*}
and the condition to eliminate one dipolar transition is 
\begin{eqnarray*}
\cos(\alpha)\mu_{13} -\sin(\alpha)\mu_{23}&=&0\,;\qquad \textrm{for}\ \ \omega_1\not\rightleftharpoons \omega_3\,,\\[2mm]
\sin(\alpha)\mu_{13} +\cos(\alpha)\mu_{23}&=&0\,;\qquad \textrm{for}\ \ \omega_2\not\rightleftharpoons \omega_3\,.
\end{eqnarray*}
For atoms with degenerate ground state one has $\omega_{21}=0$, and hence the one-body interaction term vanishes naturally. 

For atoms in the $V$-configuration one finds 
   \begin{eqnarray*}
      \op{H}_{V}^{\prime}(\alpha)&=& \Omega\,\op{a}^{\dagger}\op{a}+\omega_{1}\,\op{A}_{11}
      +\left(\omega_{2}\,\cos^{2}(\alpha)+\omega_{3}\,\sin^{2}(\alpha)\right)\,\op{A}_{22}
      +\left(\omega_{2}\,\sin^{2}(\alpha)+\omega_{3}\,\cos^{2}(\alpha)\right)\,\op{A}_{33}
      \nonumber\\
      &&+\frac{1}{2}\,\omega_{32}\,\sin(2\,\alpha)\,\left(\op{A}_{23}+\op{A}_{32}\right)
      \nonumber\\
      &&-\frac{1}{\sqrt{N_{a}}}\left(\op{a}^{\dagger}+\op{a}\right)\Big\{
      \left(\cos(\alpha)\,\mu_{12}+\sin(\alpha)\,\mu_{13}\right)
      \left(\op{A}_{12}+\op{A}_{21}\right)\nonumber\\
      &&\phantom{-\frac{1}{\sqrt{N_{a}}}\left(\op{a}^{\dagger}+\op{a}\right)}
      +\left(-\sin(\alpha)\,\mu_{12}+\cos(\alpha)\,\mu_{13}\right)
      \left(\op{A}_{13}+\op{A}_{31}\right)\Big\}\ ,
      \label{hamVrot}
   \end{eqnarray*}
similarly, to drop one dipolar transition one finds 
\begin{eqnarray*}
\cos(\alpha)\mu_{12} +\sin(\alpha)\mu_{13}&=&0\,;\qquad \textrm{for}\ \ \omega_1\not\rightleftharpoons \omega_2\,,\\[2mm]
-\sin(\alpha)\mu_{12} +\cos(\alpha)\mu_{13}&=&0\,;\qquad \textrm{for}\ \ \omega_1\not\rightleftharpoons \omega_3\,.
\end{eqnarray*}
In this case for atoms with degenerated excited states one has $\omega_{32}=0$, which eliminates in a natural way the one-body interaction term.

\section{Fidelity between neighbouring states}
\label{AppB}

Consider the overlap between a state $\psi(\mu)$ and the state $\psi(\mu+\delta\mu)$ obtained from it by applying an infinitesimal unitary transformation $\op{U}$,
   \begin{equation}
      T_{\rm rot}=\langle\psi(\mu+\delta\mu)\vert\op{U}^\dag(\mu+\delta\mu)\,\op{U}(\mu)\vert\psi(\mu)\rangle\ .
   \end{equation}
Expanding $\op{U}(\mu+\delta\mu)$ up to second order in $\delta\mu$, $T_{\rm rot}$ takes the approximate form
	\begin{eqnarray}
	   T_{\rm rot}\approx\langle\psi(\mu+\delta\mu)\vert\left(\op{U}(\mu)+\delta\mu\,\frac{d\phantom{\mu}}{d\mu}\op{U}(\mu)
	   +\frac{(\delta\mu)^{2}}{2}\frac{d^{2}\phantom{\mu}}{d\mu^{2}}\op{U}(\mu)\right)^\dag\,\op{U}(\mu)\vert\psi(\mu)\rangle\ .
	   \label{overlap}
	\end{eqnarray}
When the unitary transformation $\op{U}(\alpha)$ is of the form
   \begin{equation}
      \op{U}(\alpha)=\exp\left(-\alpha\,\op{K}\right)\ ,\quad \hbox{with}\ \op{K}^{\dagger}=-\op{K}\ ,
   \end{equation}
we have
   \begin{eqnarray}
      \frac{d\phantom{\mu}}{d\mu}\op{U}(\alpha)&=&\left(\frac{d\phantom{\alpha}}{d\alpha}\op{U}(\alpha)\right)
      \,\frac{d\alpha}{d\mu}=-\frac{d\alpha}{d\mu}\,\op{U}(\alpha)\,\op{K}\ ,\\
      \frac{d^{2}\phantom{\mu}}{d\mu^{2}}\op{U}(\alpha)&=&\left(\frac{d\alpha}{d\mu}\right)^{2}\op{U}(\alpha)\,\op{K}^{2}
      -\frac{d^{2}\alpha}{d\mu^{2}}\,\op{U}(\alpha)\,\op{K}\ .
   \end{eqnarray}
In our case, $\alpha$ depends on two parameters $\mu_{jk}$, but only a variation on one of them will be considered in the following.
Then, the squared absolute value of the overlap~(\ref{overlap}) is, up to second order in $\delta\mu$ and only considering the change
in the unitary transformation $\op{U}(\alpha)$,
   \begin{eqnarray*}
      &&F_{\rm rot}= |T_{\rm rot}|^{2}\approx |\langle\psi(\mu+\delta\mu)|\psi(\mu)\rangle|^{2}\\
      &&\quad+\delta\mu\,\frac{d\alpha}{d\mu}\,\Big\{\langle\psi(\mu+\delta\mu)|\psi(\mu)\rangle^{\ast}
      \,\langle\psi(\mu+\delta\mu)|\op{K}|\psi(\mu)\rangle\\
      &&\kern2cm-\langle\psi(\mu+\delta\mu)|\psi(\mu)\rangle
      \,\langle\psi(\mu+\delta\mu)|\op{K}|\psi(\mu)\rangle^{\ast}\Big\}\\
      &&\quad+\frac{(\delta\mu)^2}{2!}\,\Big\{\left(\frac{d\alpha}{d\mu}\right)^{2}\,\Big[\langle\psi(\mu+\delta\mu)|\psi(\mu)\rangle
      \langle\psi(\mu+\delta\mu)|\op{K}^{2}|\psi(\mu)\rangle^{\ast}\\
      &&\kern3.5cm+\langle\psi(\mu+\delta\mu)|\psi(\mu)\rangle^{\ast}
      \langle\psi(\mu+\delta\mu)|\op{K}^{2}|\psi(\mu)\rangle+|\langle\psi(\mu+\delta\mu)|\op{K}|\psi(\mu)\rangle|^{2}\Big]\\
      &&\kern2cm-\frac{d^{2}\alpha}{d\mu^{2}}\Big[\langle\psi(\mu+\delta\mu)|\psi(\mu)\rangle
      \langle\psi(\mu+\delta\mu)|\op{K}|\psi(\mu)\rangle^{\ast}\\
      &&\kern3.5cm-\langle\psi(\mu+\delta\mu)|\psi(\mu)\rangle^{\ast}
      \langle\psi(\mu+\delta\mu)|\op{K}|\psi(\mu)\rangle\Big]\Big\}\ .
   \end{eqnarray*}
Now using the fact that $\langle\psi(\mu+\delta\mu)|\psi(\mu)\rangle$, and $\langle\psi(\mu+\delta\mu)|\op{K}^{n}|\psi(\mu)\rangle$ for
$n=1,\,2$, are real numbers, the previous expression reduces to
   \begin{eqnarray}
      &&F_{\rm rot} = |T_{\rm rot}|^{2} \approx |\langle\psi(\mu+\delta\mu)|\psi(\mu)\rangle|^{2}\\
      &&\quad+(\delta\mu)^{2}\,\left(\frac{d\alpha}{d\mu}\right)^{2}\,
      \Big[\langle\psi(\mu+\delta\mu)|\psi(\mu)\rangle
      \langle\psi(\mu+\delta\mu)|\op{K}^{2}|\psi(\mu)\rangle
      +|\langle\psi(\mu+\delta\mu)|\op{K}|\psi(\mu)\rangle|^{2}\Big]\ .\nonumber
   \end{eqnarray}

We have shown numerically that, in the case considered, the minima of this last whole expression and those of only the first term in the r.h.s, are the same. The phase diagram of the ground state, therefore, for the two special rotations considered and the original Hamiltonian, remains invariant.

\vspace{2cm}

\section*{References}

\providecommand{\newblock}{}


\begin{thebibliography}{10}
\expandafter\ifx\csname url\endcsname\relax
  \def\url#1{{\tt #1}}\fi
\expandafter\ifx\csname urlprefix\endcsname\relax\def\urlprefix{URL }\fi
\providecommand{\eprint}[2][]{\url{#2}}

\bibitem{rabi37}
Rabi I~I 1937 {\em Phys. Rev.\/} {\bf 51}(8) 652--654

\bibitem{dicke54}
Dicke R~H 1954 {\em Phys. Rev.\/} {\bf 93}(1) 99--110

\bibitem{yoo85}
Yoo H~I and Eberly J~H 1985 {\em Phys. Rep.\/} {\bf 118} 239--337

\bibitem{li87}
Li X, Lin D~L and Gong C~d 1987 {\em Phys. Rev. A\/} {\bf 36}(11) 5209--5219

\bibitem{liu87}
Liu Z, Li X and Lin D~L 1987 {\em Phys. Rev. A\/} {\bf 36}(11) 5220--5225

\bibitem{lin89}
Lin D, Li X and Peng Y 1989 {\em Phys. Rev. A\/} {\bf 39} 1933

\bibitem{li89}
Li X, Lin D, George T~F and Liu Z 1989 {\em Phys. Rev. A\/} {\bf 40} 228

\bibitem{cordero13a}
Cordero S, {L\'opez-Pe\~na} R, {Casta\~nos} O and {Nahmad-Achar} E 2013 {\em
  Phys. Rev. A\/} {\bf 87}(2) 023805

\bibitem{cordero13b}
Cordero S, {Casta\~nos} O, {L\'opez-Pe\~na} R and {Nahmad-Achar} E 2013 {\em J.
  Phys. A: Math. Theor.\/} {\bf 46} 505302

\bibitem{el-wahab15}
{El-Wahab} N~A, Rady A~A, Osman A~A and Salah A 2015 {\em Eur. Phys. J. Plus\/}
  {\bf 130} 207

\bibitem{algarni22}
Algarni M, Berrada K, {Abdel-Khalek} S and Eleuch H 2022 {\em Results Phys.\/}
  {\bf 43} 106089

\bibitem{puri88}
Puri R and Bullough R 1988 {\em J. Opt. Soc. Am. B\/} {\bf 5} 2021

\bibitem{gerry90}
Gerry C~C and Eberly J~H 1990 {\em Phys. Rev. A\/} {\bf 42} 6805

\bibitem{alexanian95}
Alexanian M and Bose S~K 1995 {\em Phys. Rev. A\/} {\bf 52} 2218

\bibitem{wu96}
Wu Y 1996 {\em Phys. Rev. A\/} {\bf 54} 1586

\bibitem{zanardi06}
Zanardi P and Paunkovi\ifmmode~\acute{c}\else \'{c}\fi{} N 2006 {\em Phys. Rev.
  E\/} {\bf 74}(3) 031123

\bibitem{sachdev11}
Sachdev S 2011 {\em Quantum Phase Transitions\/} 2nd ed (Cambridge University
  Press)

\bibitem{perez-bernal08}
P\'erez-Bernal F and Iachello F 2008 {\em Phys. Rev. A\/} {\bf 77} 032115

\bibitem{cejnar21}
Cejnar P, Str\'ansk\'y P, Macek M and Kloc M 2021 {\em J. Phys. A: Math.
  Theor.\/} {\bf 54} 133001

\bibitem{miller72}
Willard~Miller J 1972 {\em Symmetry groups and their applications\/} (Academic
  Press) ISBN 978-0124974609

\bibitem{cohen20}
{Cohen-Tannoudji} C, Diu B and Lalo\"e F 2020, Weinheim {\em Quantum
  Mechanics\/} vol~II (Wiley-VCH)

\end{thebibliography}
\end{document}